\documentclass{PoS}

\newcommand{\cO}{\mathcal{O}}
\newcommand{\be}{\begin{equation}}
\newcommand{\ee}{\end{equation}}

\title{Theoretical progress on $\pi \pi$ scattering lengths and phases}

\ShortTitle{Theoretical progress on $\pi \pi$ scattering lengths and phases}

\author{\speaker{Gilberto Colangelo}\thanks{Work supported in part by the
    Swiss National Science Foundation and by the EU Contract
    No.~MRTN-CT-2006-035482, ``FLAVIAnet''.}\\ 
        Institut f\"ur Theoretische Physik\\
        Universit\"at Bern\\
        Sidlerstrasse 5, 3012 Bern\\
        E-mail: \email{gilberto@itp.unibe.ch}}

      \abstract{$\pi \pi$ scattering at low energy is sensitive to the
        structure of the QCD vacuum. I review the calculations of the $\pi
        \pi$ scattering lengths and phases, and group them in three
        cathegories: 1. those based on very general theoretical constraints
        (like dispersion relations and crossing symmetry) and
        phenomenology, 2. those which in addition make explicit use of
        chiral symmetry, 3. the first-principle ones, done with lattice
        QCD. I then compare these to the experimental results. Thanks to
        recent progress in all these and in the experimental determination
        of the scattering lengths we are improving substantially our
        knowledge of the QCD vacuum.}

\FullConference{KAON International Conference\\
		 May 21-25 2007\\
		 Laboratori Nazionali di Frascati dell'INFN, Rome, Italy}

\begin{document}

\section{Introduction}
According to the standard view, the pions become massless in the chiral
limit, because they play the role of the Goldstone bosons of spontaneous
chiral symmetry breaking. The strength of the interaction among Goldstone
bosons vanishes with the square of their momentum and with a calculable
coefficient (the inverse of the decay constant squared). If one moves
away from the chiral limit, the strength of the interaction does not
vanish when they are at rest, but for small quark masses it must be
proportional to them. The coefficient of this term is also calculable and
is given by the quark condensate (divided by the decay constant to the
fourth power). These few statements summarize what was known about $\pi
\pi$ scattering already more than 40 years ago, when Weinberg calculated the
amplitude using current algebra \cite{Weinberg:1966kf}:
\be
\label{eq:Ap2}
A(s,t,u)=\frac{s-M^2}{F^2} + \cO(p^4) \; \; ,
\ee
where $A(s,t,u)\equiv \mathcal{M}(\pi^+ \pi^- \to \pi^0 \pi^0)$ is the
isospin invariant $\pi \pi$ scattering amplitude, $M^2 \equiv B (m_u+m_d)$,
with $B=-\langle \bar q q \rangle /F^2$, is the leading term in the quark
mass expansion of the pion mass squared, $M_\pi^2=M^2+\cO(m_q^2)$,
and $F$ the pion decay constant in the chiral limit, $F_\pi= F+\cO(m_q)$.
As indicated by the $\cO(p^4)$ symbol, the result of Weinberg is the
leading term in the chiral expansion. Formula (\ref{eq:Ap2}) clearly
illustrate the importance of studying $\pi \pi$ scattering: Weinberg's
calculation heavily relies on a theoretical picture about the vacuum of
QCD. The latter is difficult to rigorously prove theoretically (indeed,
until we can prove it, this picture only has the status of a reasonable,
sound assumption), and very difficult to test experimentally. In $\pi \pi$
scattering we are now able to do it.

This took quite some efforts, however, both on the theoretical and on the
experimental side. On the theory side, one had to show that the beautiful
relation between the scattering lengths, the quark masses, the quark
condensate and the pion decay constant given by Eq.~(\ref{eq:Ap2}), valid
at leading order of the chiral expansion, does not get washed out by higher
order corrections.  Not only a next-to-leading (NLO) \cite{Gasser:1983yg},
but also a NNLO calculation \cite{Bijnens:1995yn} were necessary to reach
the required level of confidence. On the experimental side, for many years
the only known reliable method to measure the $\pi \pi$ scattering lengths
was through final state interactions in $K_{e4}$ decays
\cite{Cabibbo:1965}. It is worth stressing that this channel is quite rare
(BR$\sim 10^{-5}$), and that the effect due to the $\pi \pi$ rescattering
in the final state is rather subtle. The first measurement by the
Geneva-Saclay collaboration \cite{Rosselet:1976pu} was based on about 30000
events, came only about ten years after Weinberg prediction (with which it
disagreed!) and could only reach a precision of about 20\% on the S-wave,
isospin zero scattering length $a_0^0$.  It took more than twenty years to
see an improvement (by more than a factor of ten in the statistics) of that
experiment, by the E865 collaboration at Brookhaven \cite{Pislak:2001bf}.
Today we are in the lucky situation of having not only yet another
improvement in statistics by another experiment, NA48 (with the advantage
of having also different systematics), but also two completely
different ways to measure the $\pi \pi$ scattering lengths: the one pursued
by DIRAC  at CERN relies on the measurement of the lifetime of pionium,
whereas the one pursued again by the NA48 Collaboration, also at CERN, on a
very precise measurement of a small cusp in the spectrum of the center of
mass energy of the two neutral pions in $K^\pm \to \pi^\pm \pi^0 \pi^0$
decays.

A detailed knowledge of the $\pi \pi$ scattering amplitude is also
important for many other hadronic processes, whenever pions in the final
state play a role. Two examples of this are the determination of the
$\sigma$ resonance parameters \cite{Caprini:2005zr} (for a recent
discussion of this issue and reference to the earlier literature see
Ref.~\cite{Pennington:2007yt}), and the evaluation of the hadronic
contributions to the $g-2$ of the muon \cite{Colangelo:2006cd}. 

Since a few years, there is a new player on the $\pi \pi$ scattering arena,
lattice QCD. Various groups have calculated the isospin two, $S$ wave
scattering length $a^2_0$ in the quenched approximation and the first
calculations with dynamical fermions \cite{Yamazaki:2004qb} and reasonably
low quark masses have recently become available \cite{Beane:2005rj}. In
addition, on the lattice one can explicitly see how both the pion mass and
decay constant behave as one decreases the quark masses, and so directly
test our picture of the QCD vacuum as one approaches the chiral limit.
Using chiral perturbation theory (CHPT) one can translate this information
into values of the scattering lengths and check whether all these
informations merge into a coherent picture.

The rest of the paper is organized as follows: in the next section I will
discuss dispersion relations and in particular the Roy equations and some
phenomenological analyses based on them. These analyses rely on theory only
as far as unitarity and analyticity (and partly crossing symmetry) are
concerned, and use also data as input, but do not make any use of chiral
symmetry. In the following section I will then discuss the improvement in
precision which one obtains if one uses chiral symmetry. Section 4 gives an
overview of the recent progress in related lattice calculations. I conclude
and summarize in the final section.

\section{Theoretical calculations which do not make use of chiral symmetry}
In the early days of the study of strong interactions it was quickly
realized that perturbative methods could not be applied. In order to tackle
the problem in some more useful way, people tried to exploit the general
properties of the $S$ matrix, like analyticity and unitarity and its
possible symmetries. In $\pi \pi$ scattering this activity culminated in
the formulation of an (infinite) set of coupled dispersion relations for
all the partial waves, which incorporated analyticity and unitarity and
(partially) crossing symmetry, by S.~M.~Roy \cite{Roy}. The dispersion
relations are double subtracted, and the two subtraction constants may be
identified with the two $S$ wave scattering lengths, $a_0^0$ and $a^2_0$.
Soon after the formulation of the equations different groups started to
work on their numerical solution \cite{Pennington:1973xv,BFP} (and others
on their mathematical properties, see {\em e.g.} \cite{proofpool}, but we
will not dwell on this point here). The outcome of these analyses was that
at low energy the essential parameters are the two scattering lengths:
given some reasonable input for the imaginary parts above a certain energy
(called the matching point), the solution of the Roy equations uniquely
fixes the partial waves below this energy in terms of the two scattering
lengths.  Notice that since at low energy only the $S$ and $P$ waves have
imaginary parts significantly different from zero, the Roy equations can be
solved effectively only for these, and become a set of three coupled
integral equations. Of course the solutions do still depend on the input on
the imaginary parts of the higher waves, but these are not particularly
important and can be kept fixed (with the corresponding uncertainties 
properly taken into account). In such a setting if the uncertainty in the
input would shrink to zero, the partial waves below the matching point
would only depend on the two scattering lengths.

\begin{figure}
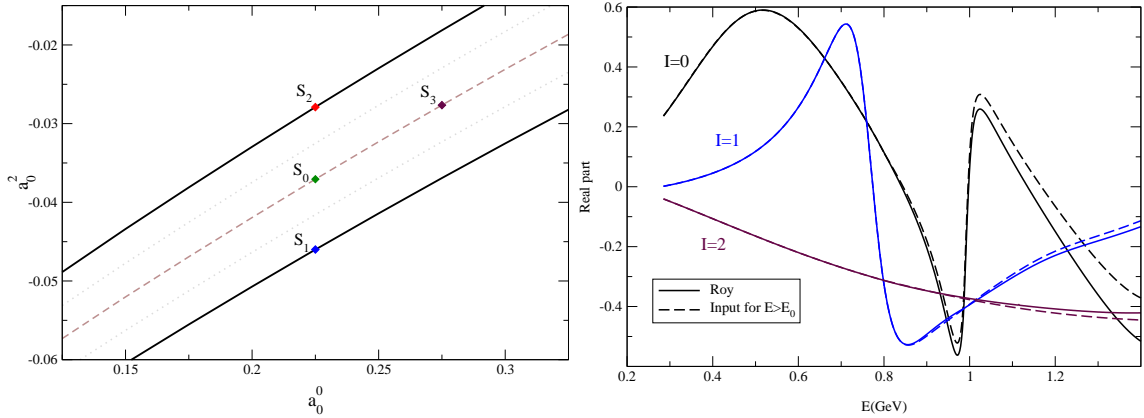

\includegraphics[width=7.5cm]{UB}
\includegraphics[width=7.5cm]{a225_example}
\caption{\label{fig:UB} Left panel: universal band. Right panel: solution
  corresponding to the point $S_0$ at the center of the universal band. The
matching point in this figure is $0.8$ GeV. Both figures are from
Ref.~\cite{ACGL}.}
\end{figure}

After about twenty years of inactivity in this field, in view of
new experiments about $\pi \pi$ scattering, the study of the
numerical solutions of Roy equations has been taken up again by a few
groups \cite{ACGL,DescotesGenon:2001tn,Kaminski:2002pe,Kaminski:2006yv}.
The results of the early analyses could be reproduced and the newly built
machineries were ready to incorporate new experimental results. The logical
flow of these analyses is as follows: take the input for the imaginary
parts above the matching point with generous uncertainties. At this stage
$a_0^0$ and $a_0^2$ are still completely free, apart from a loose
correlation which takes the form of a rather wide band, called the
universal band in the $(a_0^0, a_0^2)$ plane, see Fig.~\ref{fig:UB}, left
panel.
If one knew exactly the input for the imaginary part of the exotic $S$
wave, one could not freely choose $a_0^2$: in order to have a smooth
transition at the matching point, without unphysical cusps, $a_0^2$ has to
be appropriately tuned. In this situation, the universal band would shrink
to a line. Its width reflects the uncertainties in the experimental input
for the exotic $S$ wave \cite{losty,hoogland} (for details about this point,
cf. Ref.~\cite{ACGL}). Unfortunately it is rather unlikely that we will see
an improvement of this input -- in the foreseeable future the universal
band will stay as it is. 

To any point inside the universal band there corresponds an exact solution
of the Roy equations, as shown in Fig.\ref{fig:UB}, right panel. Such a
solution can be compared to any data set on the $\pi \pi$ scattering
amplitude below the matching point. One can then vary the two scattering
lengths and evaluate the $\chi^2$ corresponding to each point inside the
universal band. The minimum of the $\chi^2$ identifies the value of the two
scattering lengths that a certain data set prefers. Such an analysis has
been done in the early days in Refs.~\cite{Pennington:1973xv,BFP}, and more
recently repeated in Refs.~\cite{ACGL,DescotesGenon:2001tn}. The latter two
analyses agree as far as the solutions of the Roy equations are concerned
-- any difference in the conclusions arises from the use of different sets
of data, but this is much less significant than the check provided by two
completely independent analysis of the Roy equations.

This theoretical work is very important, but it is clear that by doing this
one simply relates different data sets rather than making a genuine
theoretical prediction: analyticity, unitarity and crossing symmetry do not
fully constrain the $\pi \pi$ scattering amplitude at low energy, but imply
that any measurement thereof can be translated into a measurement of the
two $S$ wave scattering lengths, and this is what such an analysis
concretely implements.

Kaminksi, Lesniak and Loiseau \cite{Kaminski:2002pe} have used the solution
of the Roy equations in order to resolve an ambiguity in the extraction of
the $\pi \pi$ scattering amplitude from $\pi N \to \pi \pi N$ data, and
have so provided another example of how useful it is to take into account
analyticity, unitarity and crossing symmetry in the data analysis. In this
manner they have also obtained ranges for the scattering lengths, which we
will compare to other analyses below.

A different approach has been followed by Pel\'aez and Yndur\'ain
\cite{Pelaez:2004vs}. They use a parametrization for each partial wave
which is simple and respects analyticity in certain low energy regions, and
fit data with these. They then check {\em a posteriori} whether forward
dispersion relations are satisfied, and use this information to improve
their fits. In a later work with Kaminski, \cite{Kaminski:2006yv} they have
concentrated on the region above the $K \bar K$ threshold and reevaluated
the dispersion relations. In comparison to other analyses, these works do
not fully exploit analyticity and crossing symmetry. Moreover, there
are some difference as far as the high-energy behaviour (described {\em \`a
  la} Regge) of the $\pi \pi$ scattering is concerned. The latter, however,
has a limited influence in the low energy region (as shown in
\cite{Caprini:2003ta} in reply to the criticism raised in
\cite{Pelaez:2003eh}). The main difference between this analysis and the
other ones concerns the behaviour of the $S0$ wave in the region between
$0.5$ GeV and the and $K \bar K$ threshold, where these authors claim that
both data as well as dispersion relations would like to have a broad
structure which they call a ``shoulder''.  As discussed by Leutwyler  
\cite{Leutwyler:2006qp}, this shoulder is in contrast with the Roy
equations.
\begin{table}
\begin{tabular}{|r|r|r|r||r|}
\hline
& \multicolumn{3}{c||}{no chiral symmetry}&with chiral symm.\\
\hline
         & DFGS \cite{DescotesGenon:2001tn} & KLL \cite{Kaminski:2002pe} &
         PY \cite{Pelaez:2004vs}& CGL \cite{CGL}\\ 
\hline
$a_0^0$  & $0.228 \pm 0.032$ & $0.224 \pm 0.013$ & $0.230 \pm 0.015$ &
$0.220 \pm 0.005$ \\
$-10 \cdot a_0^2$  & $0.382 \pm 0.038$ & $0.343 \pm 0.036$ & $0.480 \pm
0.046$ & $0.444 \pm 0.010$ \\
$(\delta_0^0-\delta_0^2)_|{_{s=M_K^2}}$ & $47.1^\circ$ &$37^\circ -
\delta_0^2(M_K^2)<49^\circ $ & $52.9^\circ \pm 1.6^\circ $ & $ 47.7^\circ
\pm 1.5^\circ$ \\ 
\hline
\end{tabular}
\caption{\label{tab:comp} Comparison of the numbers for the $S$-wave
  scattering lengths and the phase difference $\delta_0^0-\delta_0^2$ at
  $s=M_K^2$ for the theoretical analyses discussed here.}
\end{table}

A comparison of the numbers for the $S$-wave scattering lengths and of the
phase difference $\delta_0^0(M_K^2)-\delta_0^2(M_K^2)$ of the analyses just
discussed is given in Table~\ref{tab:comp}. The last column contains the
result obtained when using chiral symmetry, which will be discussed in the
next section. The main difference between the first three columns and the
last one concerns the size of the error bars -- chiral symmetry leads to a
substantial increase of the precision. The analyses which do not make use
of chiral symmetry agree among themselves as far as the central value of
$a_0^0$ is concerned. The central values of $a_0^2$ show a larger scatter,
which simply reflects the meagre experimental information about this
quantity. Finally, the last row shows that the analysis of Pel\'aez and
Yndur\'ain has a higher $\delta_0^0$ phase around the $K$ mass. Direct
extraction of the phase difference $\delta_0^0(M_K^2)-\delta_0^2(M_K^2)$
from $K \to 2 \pi$ decays actually indicate an even larger value. After
applying isospin breaking corrections, Cirigliano {\em et al.}
\cite{Cirigliano:2003gt} find $(\delta_0^0-\delta_0^2)_|{_{s=M_K^2}} =
(60.8 \pm 4)^\circ$. The recent update of the measurement of $\Gamma(K_S
\to \pi^+ \pi^- (\gamma))/\Gamma(K_S \to \pi^0 \pi^0)$ by KLOE
\cite{Ambrosino:2006sh} pushes this number down by about three 
degrees. Notice that the isospin breaking correction is in this case very
large, about $12^\circ$ \cite{Cirigliano:2003gt} -- its evaluation is far
from straightforward, and despite the careful analysis of Cirigliano {\em
  et al.} the outcome is puzzling. Understanding the clash between the
value extracted from $K \to 2\pi$ and all other experimental and
theoretical analyses remains an open, very important problem\footnote{In a
  recent paper, \cite{Yndurain:2007qm} it has been proposed to determine
  the scattering lengths from a simultaneous fit to the $K_{e4}$ data and
  to the phase extracted from $K \to 2 \pi$. The large uncertainties in the
  latter extraction and the puzzling result would rather suggest not to use
  this experimental information in any fit.}.

In all these analyses, input on the $\pi \pi$ scattering amplitude at
intermediate energies is essential. This is obtained from different
experiments by also using some convenient parametrizations which
incorporate at least partially the properties of analyticity and
unitarity. Until some years ago this input came mostly from $\pi N \to \pi
\pi N$ scattering experiments \cite{hyams,Protopopescu,Grayer,EM}. More
recently, production of two (or more) pions in hadronic decays of heavier
states have also become important, in particular for what concerns the
resonances in the region around $1$ GeV, like the $f_0(980)$, which has
been studied in detail at BES \cite{BES f0(980)} and at KLOE
\cite{Ambrosino:2006hb}, see also
\cite{Achasov:2005hm,Isidori:2006we,Bugg:2006sr}.

\section{Theoretical calculations which rely on chiral symmetry}
\begin{figure}[thb]
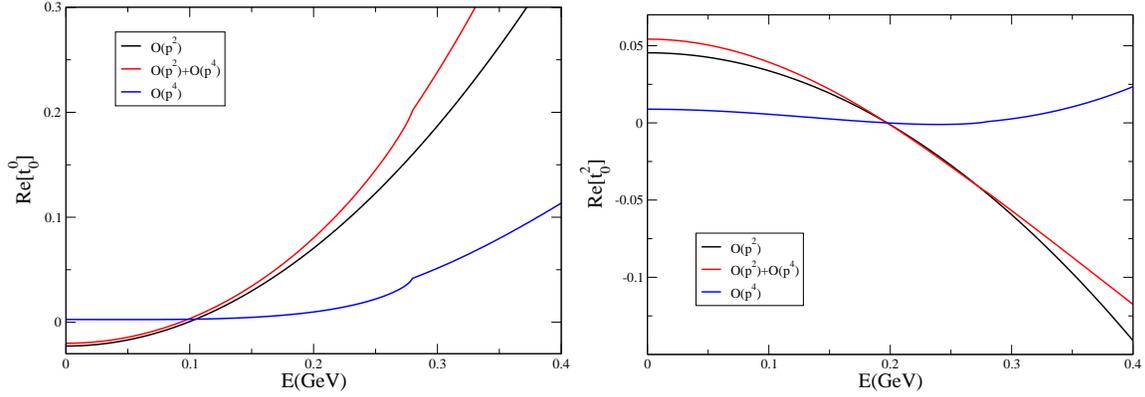

\includegraphics[width=7.5cm]{t00_24}
\includegraphics[width=7.5cm]{t20_24}
\caption{\label{fig:rescatt} Behaviour of the $I=0$ (left panel) and $I=2$
  $S$ waves in the region around threshold.}
\end{figure}
The Roy equations allow to translate any experimental information on $\pi
\pi$ scattering at low energy into information on the scattering
lengths. But one can also reverse the argument: if one has a
theory which predicts the scattering lengths, one can combine this with the
Roy equations and extend the prediction to the whole region below 1 GeV.
As we have mentioned in the introduction, chiral symmetry does lead to
predictions about the scattering lengths. The leading order calculation of
Weinberg gives (after one substitutes $M^2$ and $F$ with $M_\pi^2$ and
$F_\pi$, respectively)
\be
a_0^0= \frac{7 M_\pi^2}{32 \pi F_\pi^2}=0.16 \; , \qquad
a_0^2=-\frac{M_\pi^2}{16 \pi F_\pi^2}=-0.045 \; \; .
\ee
The theoretical uncertainties of this prediction cannot be estimated until
one evaluates the next term in the series. This was done by Gasser and
Leutwyler in 1984 \cite{Gasser:1983yg}, and some ten years later even the
correction one order higher has been evaluated
\cite{Bijnens:1995yn}. Numerically the series behaves as follows:
\be
a_0^0\stackrel{p^2}{=}0.156 \stackrel{p^4}{\to} 0.200 \stackrel{p^6}{\to} 0.216 
\; , \qquad a_0^2\stackrel{p^2}{=}-0.0454 \stackrel{p^4}{\to} -0.0445
\stackrel{p^6}{\to} -0.0445 \; \; ,
\ee
and shows a rather slow convergence in the $I=0$ channel. The reason for
this slow convergence is understood and has to do with a rather hefty chiral
log:
\be
a_0^0=\frac{7M_\pi^2}{32\pi F_\pi^2}\left[1+\frac{9}{2} \ell_\chi +
  \ldots \right] \; , \qquad
a_0^2=-\frac{M_\pi^2}{16 \pi F_\pi^2}\left[1-\frac{3}{2} \ell_\chi +
  \ldots \right]
\ee
where $\ell_\chi=\frac{M_\pi^2}{16 \pi^2 F_\pi^2} \ln \frac{\mu^2}{M_\pi^2}$
which in turn is due to a strong rescattering of pions in this
channel. The large corrections are unitarity effects, as it is also very
well illustrated in Fig.~\ref{fig:rescatt}: the
correction of order $p^4$ shows a strong curvature below threshold and a
sizeable cusp at threshold in the $I=0$ channel, but is very flat below and
does not have a visible cusp at threshold in the $I=2$ channel. Around
$s=0$, however, both corrections are very small.

This observation is crucial if one wants to combine the chiral prediction
and the dispersive analysis: choosing the two scattering lengths as
subtraction constants is convenient for discussing the physics, but it is
not the only possible choice. One could as well subtract the amplitudes in
the region below threshold. Doing this has the advantage that if one uses
the chiral input there, this converges much better and gives stability to
the whole machinery. Indeed, after subtracting below threshold, and
evaluating the scattering lengths with the help of the Roy equations
\cite{CGL}, the behaviour of the series improves drastically:
\be
a_0^0\stackrel{p^2}{=}0.197 \stackrel{p^4}{\to} 0.2195 \stackrel{p^6}{\to}
0.220  
\qquad a_0^2\stackrel{p^2}{=}-0.0402 \stackrel{p^4}{\to} -0.0446
\stackrel{p^4}{\to} -0.0444 \; \; .
\ee
Within this framework, the evaluation of the uncertainty can be done
reliably, and gives
\begin{eqnarray}
a_0^0&=&\; \; \, 0.220\pm0.001+0.027 \Delta_{r^2}-0.0017 \Delta \ell_3
\nonumber \\
10\cdot a_0^2&=&-0.444 \pm0.003-0.04 \Delta_{r^2}-0.004 \Delta \ell_3
\end{eqnarray}
where $\bar \ell_3=2.9 + \Delta \ell_3$, is the
low-energy constant which determines
the next-to-leading quark mass dependence of the pion mass:
\be
M_\pi^2=M^2\left(1- \frac{M^2}{32 \pi^2 F^2} \bar \ell_3+O(M^4)  \right)
\ee
and $\Delta_{r^2}$ is the relative uncertainty in the scalar radius of
the pion, $\langle r^2 \rangle_s = 0.61 \,\mathrm{fm}^2 \cdot
(1+\Delta_{r^2})$. The scalar radius strongly depends on the low energy
constant $\bar \ell_4$, which determines the leading quark mass dependence
of the pion decay constant: 
\be
F_\pi=F\left(1+ \frac{M^2}{16 \pi^2 F^2} \bar \ell_4+O(M^4)  \right) \; .
\ee
Adding errors in quadrature and using the estimates
$\Delta_{r^2}=6.5\%$ (which, together with the central value quoted above,
corresponds to $\bar \ell_4 = 4.4 \pm 0.2$), $\Delta \ell_3=2.4$ yields
\cite{CGL}   
\be
a_0^0=0.220\pm0.005 \; , \qquad 
a_0^2=-0.0444 \pm0.001 \; , \qquad 
a_0^0-a_0^2= \; \; \, 0.265\pm0.004 \; .
\label{eq:a0a2}
\ee
\begin{figure}[t]
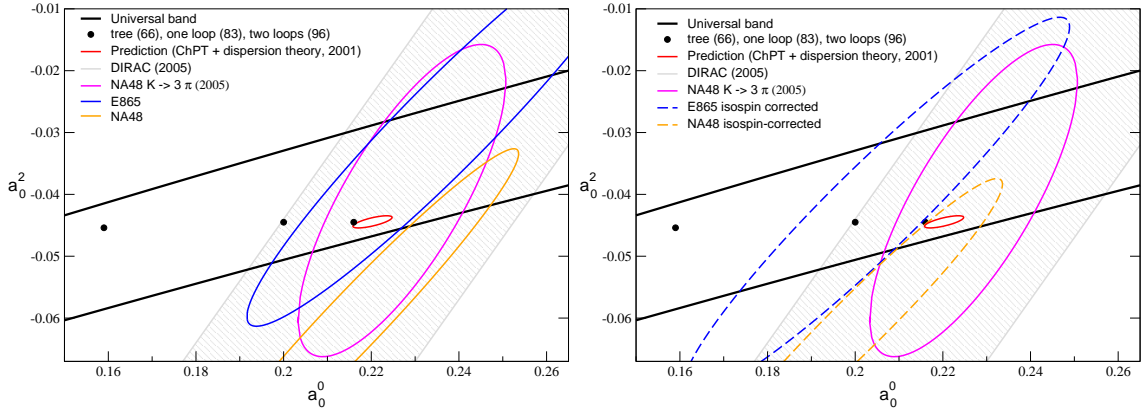

\includegraphics[width=7.5cm]{all+Ke4_NA48-unc}
\includegraphics[width=7.5cm]{all+Ke4_NA48-corr}
\caption{\label{fig:a0a2}
  Comparison of the theoretical predictions of the scattering
  lengths and their measurements. On the left the blue and orange ellipses
  are obtained from the uncorrected $K_{e4}$ and on the right those after
  isospin corrections (cf. Ref.~\cite{Gasser})}
\end{figure}

The experimental determinations of the scattering lengths have been amply
discussed at this conference \cite{exp}. The comparison between the
experimental numbers and the theoretical predictions is shown in
Fig.~\ref{fig:a0a2}. On the left panel the ellipses corresponding to the
$K_{e4}$ data sets have been obtained with the raw data, whereas  on the
right panel the isospin breaking correction to the phase as extracted from
$K_{e4}$ data, which has been evaluated and discussed in
Ref.~\cite{Gasser}, has been applied. The figure shows that the latter
isospin breaking correction is important at the current level of precision
of the experiments. The disagreement at the level of 1.5 $\sigma$'s between
the recent NA48 determination and the theoretical prediction disappears once
this correction is taken into account. On the other hand, the perfect
agreement seen on the left panel between the E865 determination and the
theoretical prediction, becomes marginal, at the level of one
sigma. In either case there is some tension between the E865 and the NA48
$K_{e4}$ data, which should be better understood This issue is discussed
in detail in the contribution of Bloch-Devaux \cite{exp}.

\section{Lattice calculations}
\begin{figure}
\begin{center}
\includegraphics[width=12cm]{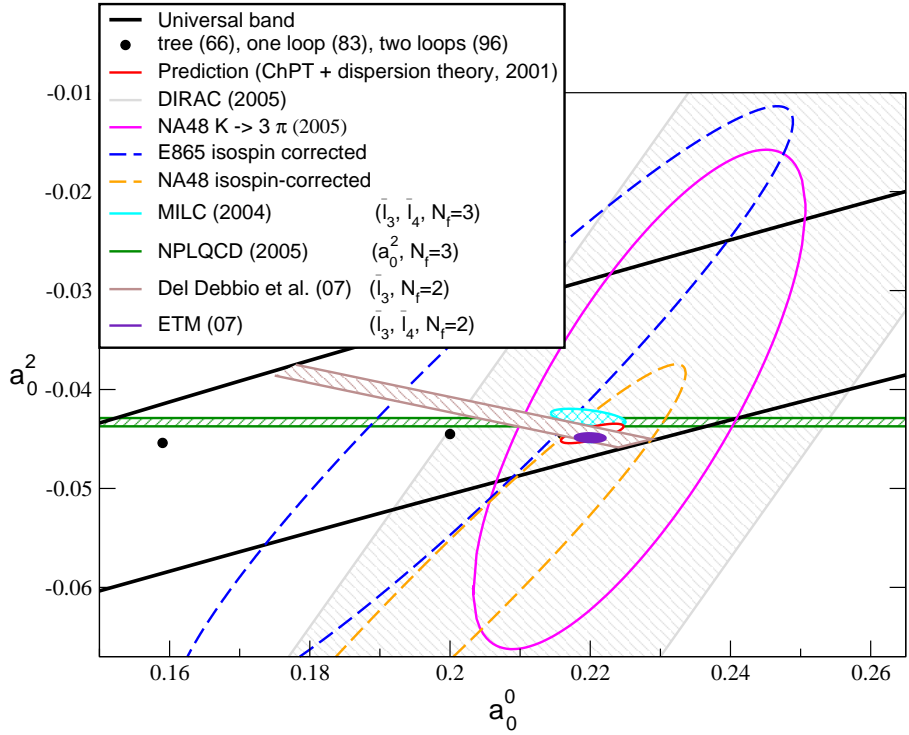}
\caption{\label{fig:a0a2lat} Same as Fig.~\protect\ref{fig:a0a2}, right panel, but
  including also lattice results.}
\end{center}
\end{figure}

Lattice calculations relevant for $\pi \pi$ scattering can be grouped into
two classes: those which determine the quark mass dependence of $M_\pi$ and
$F_\pi$ and thereby determine the constants $\bar \ell_3$ and $\bar
\ell_4$; and those which determine directly the scattering lengths. 
There are only two calculations of $a^2_0$ available until now with
dynamical fermions, one of them is performed on a background containing
only two dynamical quarks \cite{Yamazaki:2004qb}, while the more recent one
by the NPLQCD collaboration is performed on a background of three flavours
of staggered quarks \cite{Beane:2005rj} (on configurations generated by the
MILC collaboration and made openly accessible).
The latter calculation was done with a rather low pion mass, reaching
values just below 300 MeV, such that an extrapolation down to physical pion
masses becomes reliable. Their latest result reads
\be
a_0^2=-0.04330 \pm 0.00042
\ee
in excellent agreement with the chiral prediction (\ref{eq:a0a2}), as it is
also seen on Fig.~\ref{fig:a0a2lat}.
The CP-PACS calculation, on the other hand has
been made for a value of the pion mass above 500 MeV, where contact with
chiral perturbation theory, or an extrapolation to the physical value of
the pion mass can hardly be possible. For the earlier literature on the
subject, in particular on the quenched calculations, we refer the reader to
Ref.~\cite{Beane:2005rj}.

The determination of the quark mass dependence of the pion mass and decay
constant with dynamical fermions and for low pion masses has been performed
by several groups. Published results are available from the MILC
collaboration \cite{Aubin:2004fs}, from Del Debbio {\em et al.} \cite{Del
  Debbio:2006cn} and from the ETM collaboration \cite{Boucaud:2007uk}. Only
the first calculation has been done with a background of three dynamical
flavours (of staggered quarks and employing the fourth root trick), whereas
the last two have two light quarks in the sea (with a Wilson and twisted
mass formulation, respectively). A summary of their numerical results is
given in Table~\ref{tab:latt}. 
\begin{table}[t]
\begin{center}
\begin{tabular}{|r|r|r|r|}
\hline
 & MILC \cite{Aubin:2004fs} & Del Debbio {\em et al.} \cite{Del
   Debbio:2006cn} & ETM \cite{Boucaud:2007uk}\\ 
\hline
$\bar \ell_3$ & $0.6 \pm 1.2$ & $3.5 \pm 0.5 \pm 0.1$ & $3.65 \pm 0.12$ \\
$\bar \ell_4$ & $3.9 \pm 0.5$ & & $4.52 \pm 0.06$ \\
\hline
\end{tabular}
\caption{\label{tab:latt} Lattice determinations of the low energy
  constants $\bar \ell_3$ and $\bar \ell_4$.}
\end{center}
\end{table}
The agreement with the phenomenological
estimates is remarkable. It is important to stress the improvement in
precision that the lattice approach offers for these constants, in
particular for $\bar \ell_3$. In the long run the lattice method is without
competition for this particular constant, and will compete with the
phenomenological determination of others (notice that the errors given by
the ETM collaboration do not include systematic effects -- these have been
estimated by the other two groups).

\section{Summary and conclusions}
The $\pi \pi$ scattering amplitude at low energy is one of the rare
physical quantities that we can calculate with a high precision and that at
the same time can be measured with a comparable precision. In addition, the
comparison is very instructive, because we can relate the theoretical
predictions made within the effective field theory framework to properties
of the vacuum state of QCD. The recent progress in lattice calculations
makes this issue even more interesting, because it allows us to compare
experimental numbers directly to the result of first principle
calculations, which only take as input the Lagrangian of QCD. All this is
very well represented by Fig.~\ref{fig:a0a2lat}, which shows (albeit in
rather compressed form) the convergence of all these different
informations.  The figure also shows that we have room for improvement,
especially on the experimental side, and possibly for surprises. It will be
interesting to see how this picture will look like at the next Kaon
conference.

\section*{Acknowledgments}
It is a pleasure to thank the organizers for the invitation and the
excellent organization of a very exciting conference. I thank Irinel
Caprini, J\"urg Gasser and Heiri Leutwyler for a careful reading of the
manuscript and a longstanding and pleasant collaboration on many of the
issues discussed here. I thank Vincenzo Cirigliano for providing
information about the extraction of the phase difference from $K \to 2 \pi$
decays, and J.R.~Pel\'aez and F.J.~Yndur\'ain for comments on the
manuscript.

\end{document}